\documentclass{sig-alternate-05-2015}

\usepackage{amsmath}
\usepackage{amsfonts}
\usepackage{amssymb}
\usepackage{bbm}
\usepackage{balance} 
\usepackage{booktabs}
\usepackage{cite}
\usepackage{capt-of}
\usepackage{color}
\usepackage{mdwlist}
\usepackage{graphicx}

\usepackage{enumitem}
\usepackage{mathtools}
\usepackage{multirow}
\usepackage{rotating}
\usepackage{subfigure}
\usepackage{times}
\usepackage{tikz}
\usepackage{url}
\usepackage{xspace}

\begin{document}

\CopyrightYear{2016} 
\setcopyright{acmcopyright}
\conferenceinfo{WebSci '16,}{May 22-25, 2016, Hannover, Germany}
\isbn{978-1-4503-4208-7/16/05}\acmPrice{\$15.00}
\doi{http://dx.doi.org/10.1145/2908131.2908159}

\title{Understanding Video-Ad Consumption on YouTube}
\subtitle{A Measurement Study on User Behavior, Popularity, and Content Properties} 

\newcommand\Mark[1]{\textsuperscript{#1}}
\numberofauthors{1}
\author{
\alignauthor Mariana Arantes\Mark{1} \quad Flavio Figueiredo\Mark{2} \quad Jussara M. Almeida\Mark{1}\\
\affaddr{\Mark{1}Universidade Federal de Minas Gerais - Brazil}\\ 
\affaddr{\Mark{2}IBM Research - Brazil} 
\email{\Mark{1}\{mariana.arantes, jussara\}@dcc.ufmg.br \quad \Mark{2}flaviovdf@br.ibm.com} 
}

\maketitle

\begin{abstract}
Faced with the challenge of attracting user attention and revenue, social media websites have turned to video advertisements (video- ads). While in traditional media the video-ad market is mostly based on an interaction between content providers and marketers, the use of video-ads in social media has enabled a more complex interaction, that also includes content creator and viewer preferences. To better understand this novel setting, we present the first data-driven analysis of video-ad exhibitions on YouTube.

\end{abstract}



\section{Introduction}
\label{sec:intro}
From search engines to social media applications, advertising has become an ubiquitous commerce on the most popular Internet websites. In these websites, users are given access to a wide range of content (e.g., YouTube videos), whereas the content providers that maintain the application exploit user behavior and content data to create online ad-auctions to earn profits~\cite{Abraham2013,Amarie2014a,Amarie2014,Carrascosa2014,Carrascosa2013,Farahat2012,Ghose2010,Ghosh2015,Gill2013,Krishnan2013,Lacerda2006,Liu2014}. 


Social media applications, in particular, allowed for a novel advertisement market. In these applications, any user can take the role of content creator, viewer or marketer. In contrast, in print or television advertising, the creation and selection of advertisement placements is done by select individuals. Taking YouTube as an example, any user can create videos and market them as advertisements to be streamed to other users. This market allows any user to profit from ads, be it as a content producer that receives monetary shares when ads are streamed before their videos, or even as viewers that can gain from well placed ads that meet their preferences. 

We here present a measurement study of how video-ads present on the currently most popular video streaming application, YouTube, are consumed by users. Video-ads, i.e., advertisements presented to the user in the form of  videos, is rising as one of the most important (in terms of investments and profits) means of online publicity~\cite{Variety2015,Variety2014,Wall-Street-Journal2014}. Thus, to better understand the video-ads market on YouTube, in this paper we study three research questions (RQs): 

\noindent {\bf RQ1: How do users consume video-ads?}
 YouTube allows users   to skip watching the ad entirely, usually after an initial exhibition period (e.g., 5 seconds), jumping directly to the requested video. In RQ1, we  analyze how users consume video-ads, focusing on their ``skipping'' behavior. Specifically, we characterize user behavior
 as to whether they tend to consume video-ads in full or  skip them, and the fraction of the ad exhibited to the user prior to the skipping.  By tackling RQ1, we aim at drawing insights into how users often respond to video-ad exhibitions,  and how effective these ads are in terms of drawing and {\it keeping} user attention, particularly if compared to other forms of online advertising \cite{Schneider2009,Krishnan2013}.

\noindent {\bf RQ2: How does video-ad popularity evolve over time?} We aim at understanding the properties of video-ad popularity by analyzing the distributions of the {\it number of views} and {\it exposure time}. Whereas the former captures the amount of accesses to each video-ad, the latter captures the amount of time that users were exposed to its content.
While measuring the effectiveness of ads is a controversial research issue, both number of views and exposure time have been used as proxies of success of ad campaigns~\cite{Farahat2012,Dreze2004,Ghose2010,Manchanda2006}. We  look into how bursty the popularity evolution of video-ads is, the time it takes for a video-ad to peak in popularity, and the different profiles of video-ad popularity evolution. Thus, in RQ2, we deepen our analyses of the effectiveness of video-ads on YouTube,  correlating the profile of popularity evolution followed by a video-ad with its ultimate success (in terms of popularity).

\noindent {\bf RQ3: What are the relationships (if any) between a video-ad and the video-contents with which it is associated?}  In ad-auctions, a video-ad is paired with a piece of content (a YouTube video in our case, or simply a video-content)  to be displayed to the user. Our aims in RQ3 are twofold. First, we analyze whether more popular video-ads tend to be paired with videos that are also very popular. Secondly, we assess the extent to which video-ads that are paired with more similar content have a tendency to be more effective (popular), thus  uncovering evidence of whether contextual advertising~\cite{Lacerda2006} increases the effectiveness of video-ads on YouTube. 


The main contribution of this study is to provide an in-depth view of different properties of video-ads on YouTube. Our findings offer a {\it novel}, {\it broad} and {\it timely} look into the ecosystem of video advertisements, drawing valuable insights that motivate the design of more cost-effective strategies to make online video-ads potentially more profitable. Such insights should be of interest to content producers, content providers and marketers, who financially benefit from the success of ad campaigns.  Our findings are also of interest to YouTube users in general since they are subject to video-ads.

\section{Related Work}
\label{sec:related}
In contrast to the large amount of research that  has been done in online advertising in general~\cite{Abraham2013,Amarie2014a,Amarie2014,Carrascosa2014,Carrascosa2013,Farahat2012,Ghose2010,Ghosh2015,Gill2013,Krishnan2013,Lacerda2006,Liu2014},  video-ads have only been studied very recently~\cite{Krishnan2013,Amarie2014,Amarie2014a}. 

In particular, Amarie {\it et al.}~\cite{Amarie2014,Amarie2014a} focused on caching strategies for mobile advertisements in video form. In order to motivate such strategies, the authors characterized the following properties of a small sample (458) of video-ads shown in mobile devices: size (in bytes), duration, category and time of day when the video-ad is streamed. This work is complementary to our present effort. While the authors did look into some properties of video-ad popularity, their study is focused on a small sample of ads shown in mobile devices only. Moreover,  they did not study video-ad popularity evolution, content properties of video-ad to video-content pairings, and user consumption behavior, as we do here.

Stepping away from  social media applications, Krishnan {\it et al.}~\cite{Krishnan2013} characterized a large sample of video-ads 
streamed from professional content websites (e.g., NBC, CBS, CNN, Hulu, Fox News etc.)  using Akamai's content distribution network (CDN). 
One of the results reported by the authors is that video-ads have completion rates (fraction of ads that are streamed in their full length to the users) ranging from 44\%, when shown after the video-content, to 96\%, when shown in the middle of video-content. They also showed that longer video-contents have higher video-ad completion rates. 
However, the applications analyzed by the authors did not allow users to skip video-ad exhibition and jump to the video-content: users had to abandon watching the video-content altogether so as to stop watching the video-ad.  YouTube users, on the other hand, are allowed to skip the video-ad, jumping directly to the video-content, typically after an initial exhibition period. Thus, unlike in \cite{Krishnan2013}, we here study user consumption behavior in a broader sense, by analyzing the fraction of time  users were exposed to the video-ad before skipping it.  We also tackle novel aspects of video-ad consumption which were not discussed in \cite{Krishnan2013}, notably popularity evolution and video-content to video-ad relationships. 


Finally, various previous efforts have looked into different properties of user-behavior from HTTP requests~\cite{Schneider2009}, the use of social media for advertising~\cite{Carrascosa2014,Carrascosa2013}, properties of search engine ads~\cite{Farahat2012,Ghose2010,Ghosh2015}, ad-auctions in general~\cite{Gill2013,Liu2014}, as well as network characteristics of video streams~\cite{Rao2011,Gill2007}. Our work is also orthogonal to all these prior studies as it focuses  on novel aspects of video-ad consumption. In the next section, we detail our datasets. 

\section{Data Collection and Cleaning}
\label{sec:data}

We start this section by introducing some concepts  that are used throughout this paper.  We use the term {\bf video-ad} to refer to an advertisement presented to the user in the form of a video.  Each such video-ad is associated with one (or multiple) pieces of content (videos on YouTube), referred to as  {\bf video-content}. 
The association of a video-ad to a video-content is referred to as a {\bf pairing}. A pairing of a video-ad with a video-content is done in  {\it real time}, that is, at the time a user requests the video-content. Thus, multiple video-ads (as  no video-ad at all) may be paired with the same video-content as response to different requests to the same content. A video-ad {\bf exhibition} refers to the (partial or complete) streaming of the ad while paired with a given video-content, and the time period during which a particular user was exposed to a video-ad exhibition is referred to as  {\bf exhibition time}.  Finally, the {\bf exposure time} of a video-ad refers to the total amount of time (all) users dedicated to streaming the given video-ad (i.e, total exhibition time).

 We also note that each video-ad is a video by itself on YouTube. As such, it has a system id and a webpage containing the video, its  associated metadata and statistics kept by the system, as further discussed below. Moreover, each video-ad may also be requested directly, without  being paired with other videos.  Thus, in our study, a video-ad is  ultimately any video that is used as an advertisement by being paired with other video-contents in the system. 

In order to provide answers to our three research questions, we combined data from two rich and complementary sources. Initially, we collected HTTP requests  from a university campus network  to analyze user behavior when exposed to video-ads. From these requests, we filtered every video-ad to video-content pairings (both uniquely identified by system ids) that occur when video advertisements are displayed in YouTube videos. This dataset  was combined with the public information available from the YouTube's API\footnote{\url{http://developers.google.com/youtube/}} and statistics provided on the HTML content of the video page.
Such information allowed us to analyze global properties of video-ad consumption, while still focusing on the same video-ad and video-content pairings present in our HTTP requests.



\subsection{Capturing User Consumption Behavior} \label{sec:localtrace}

In order to capture user behavior in terms of how they consume video-ads on YouTube, we relied on logs of HTTP requests originating from the campus network of a major Brazilian university, with a population (including students, faculty and staff) of over 57 thousand people. Specifically, we captured the outgoing/incoming HTTP traffic from the local campus network  using 
TSTAT~\cite{Finamore2011}. The tool provides us  the headers,   originating IP addresses, and timestamps of each request/response pair. Our goal was then to extract from these requests each video-ad to video-content pairing, as well as the  exhibition time of the video-ad in each such pairing.  This was a challenging task, as, in the absence of prior studies of video-ad requests to YouTube, we did not know how to identify neither the pairings nor the exhibition times  in the traffic log. 

Thus, we started by first manually identifying different request patterns for video-ads. We did so by browsing different YouTube videos and using network analysis tools provided  by modern browsers (e.g., Firefox and Google Chrome) to assist in our investigation. 
We were able to identify request patterns for video-ads exhibited on: (1) the YouTube website; (2) embedded videos on different websites\footnote{We also attempted to identify video-ad requests from mobile devices. However, due to the different YouTube streaming applications (e.g., Android and IOS), as well as different mobile browser request patterns, we were unable to identify a representative set of requests to cover the various means of exhibiting YouTube video-ads on mobile devices. We leave this task for future work.}. These requests contain the unique YouTube identifiers of both video-ad  and  video-content, as exemplified below:


\vspace{-0.1cm}
\begin{scriptsize}
\begin{verbatim}
(1) ...youtube.com/api/stats/ads? 
            ad_v=WVgYOaERNj4& 
            content_v=-faTXv3Frc0&...  
(2) ...youtube.com/yva_video?
            video_id=WVgYOaERNj4&
            content_v=-faTXv3Frc0&...
\end{verbatim}
\end{scriptsize}
\vspace{-0.1cm}

\noindent   In requests  to the YouTube's website (example (1)), the unique id of the video-ad  is captured by the \texttt{ad\_v} parameter.  In requests for  embedded video (2), it is identified by the \texttt{video\_id} parameter. In both cases, the video-content id is captured by the  \texttt{content\_v} parameter. Using only these requests, it is possible to identify all ad to content pairings that occurred inside the campus network, but {\it not} the video-ads'  exhibition times. In order to capture this metric, we identified two other HTTP requests that are triggered when: (3) the video-ad is exhibited in full to the user; (4) the  video-ad is exhibited only partially as the user skips it after a certain initial period of streaming. Examples of these two  request types are shown below:

\vspace{-0.1cm}
\begin{scriptsize}
\begin{verbatim}
(3) ...doubleclick.net/pagead/conversion
            label=videoplaytime100&...
(4) ...doubleclick.net/pagead/conversion
            label=videoskipped&
            len=30&
            skip=6&...
\end{verbatim}
\end{scriptsize}
\vspace{-0.1cm}

\noindent In (3), the video-ad  was streamed  until completion (as identified by \texttt{videoplaytime100}), while in (4) the user skipped  the video-ad exhibition after 6 seconds (as identified by the \texttt{skip} parameter). Notice that neither  request contains any parameter that can be used to identify the ids of the video-content and the video-ad. 
\begin{figure}[t]
\centering
\includegraphics[width=1.0\linewidth]{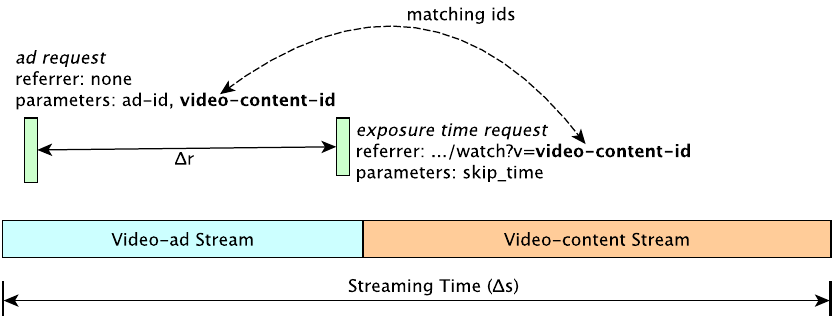}
\caption{Matching video-ad ids to video-content ids to identify ad to content pairings.}
\label{fig:match}
\vspace{-0.2cm}
\end{figure}

In order to match the {\it video-ad requests} (1-2) to the {\it exhibition time requests} (3-4), we made use of the HTTP referrer field, which captures the URL from which the user originated the HTTP request. All exhibition time requests have the page of a YouTube video-content as referrer, regardless of whether the request was triggered from YouTube's website or from an embedded video\footnote{In the cases of embedded videos, it would be expected that  the referrer field in the requests in examples (3) and (4) would be equal to the URL that embedded the video.  However, we found that the referrer is always a YouTube video page given that the video-player is actually hosted on \texttt{youtube.com}.}. Making use of the referrer field, we were able to match the video-ad requests to the  exhibition time requests using the following simple heuristic, which is illustrated in Figure~\ref{fig:match}.

Let us define $|\Delta_r|$ as the shortest {\it absolute}\footnote{We use absolute values of $\Delta_r$ as there is no guarantee that the video-ad request will precede the exhibition time request.} time interval between a video-ad request and an exhibition time request that meets the following criteria: (a) both requests originated from the same IP address; (b) the video-content id on the referrer of the exhibition time request matches the \texttt{content\_v} parameter on the video-ad request. Also, let us define $\Delta_s$ as the time the user spends  streaming both the video-ad and the video-content. We consider that a successful match occurs between a video-ad and an exhibition time request that meet the above criteria whenever 
$|\Delta_r| < \Delta_s$. 
 Otherwise, we discard the video-ad request as an unsuccessful match. 

The above heuristic would be sufficient if {\it network address translation (NAT)} was {\it not} present in the campus network, which we cannot guarantee. Due to NAT, 
multiple exposure time requests from the same IP may have the same video-ad request as a candidate match (i.e., with the shortest  $|\Delta_r|$).
We call this case a conflict.
To deal with these conflicting matches, we initially consider as  successful  the  match with the shortest $|\Delta_r|$ out of all matches in conflict. We then remove the matched video-ad and exhibition time requests from the HTTP trace, updating $|\Delta_r|$ for all other conflicts\footnote{In practice, the HTTP trace is not altered, the whole process can be done in linear time by keeping track of conflicts in dictionaries.}. This is done by considering the next video-ad request with the shortest $|\Delta_r|$ as a match for the remaining conflicted exposure time requests. The process is  repeated for every conflict.

$|\Delta_r|$ can be computed directly from the timestamps of the HTTP requests, as shown in Figure \ref{fig:match}. $\Delta_s$ was approximated by the sum of: (1) the video-content duration
(obtained from the API,  as discussed below) 
and (2) the value of the \texttt{skip} parameter of the exhibition time request (for partial exhibitions of the video-ad) or the video-ad duration (for full exhibitions).  Video-content and video-ad durations were obtained from the API (next section).
Whenever the video-content or video-ad was not available in the API,  we used the average value of the respective duration.

It is important to point out that, while the use of the total duration of the video-content will fail to capture the behavior of users that abandon watching the content, our goal with this heuristic is to {\it simply} match video-content to video-ad pairs and not to capture the amount of time the video-content was streamed. One possible issue that may rise with the use of the total duration is a {\it false positive} on our matching heuristic. However, such cases are similar to the above described conflicts, where we may falsely match a video-content to a video-ad. Nevertheless, this situation is also dealt with our conflict resolution strategy, given that we keep the match closest to when the video-content began streaming. 

In our study we analyze the behavior of users from an aggregated level. That is, due to privacy ethics and NAT, the IP addresses (which are anonymized in our dataset)  are used in our matching heuristic, they are not used in any of our analyses. Moreover, because of the possible presence of NAT, we only analyze user behavior in terms of individual video-ad exhibitions. One limitation of our dataset is that we do not have demographical data of every member of the academic population, and thus we are unable to study targeted ads to individual users. However, our goal with this study is to uncover properties on the skipping behavior of users, popularity properties of video-ads and study contextual advertisements. We leave the task of analyzing personalized ads as future work. Nevertheless, we can state that  based on the public campus census, the university is attended by students from all over the country, most of them are in the 20-24 age range and there is a roughly equal number of men and women.

\subsection{Capturing Global Properties of Video-Ads} \label{sec:globaltrace}

\begin{figure}[t]
\centering
\includegraphics[width=1.0\linewidth]{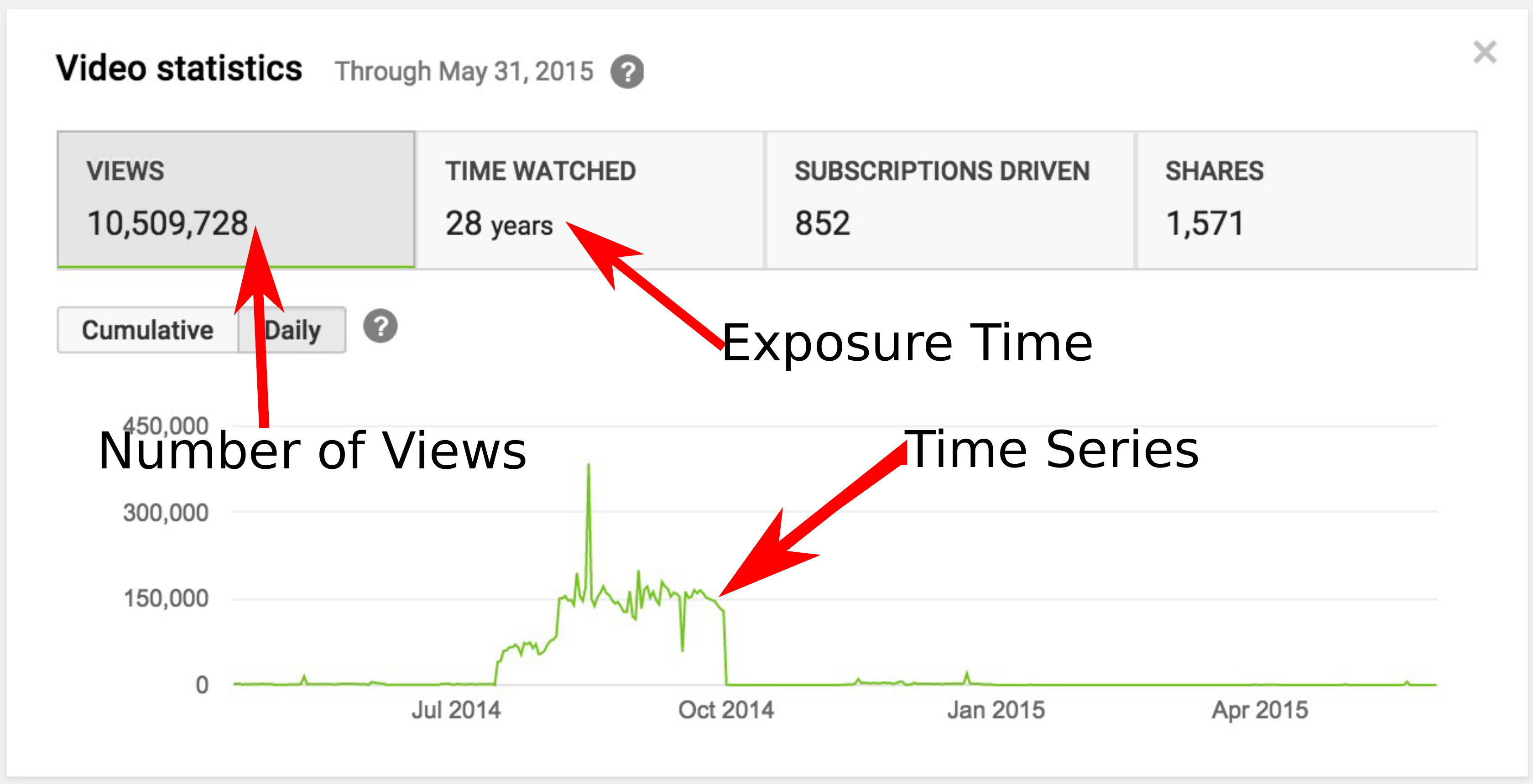}
\caption{Public statistics data provided by YouTube.}
\label{fig:stats}
\vspace{-0.2cm}
\end{figure}

We crawled the public API\footnote{\url{http://developers.google.com/youtube/}} information provided by YouTube for each unique id of
 video-content and video-ad present in our HTTP request dataset.  Specifically, for each video-content/video-ad, we collected the following metadata: {\it upload time},  {\it duration} (in seconds),  {\it title},    {\it description}, 
  {\it category}, and list of {\it topics}. Title and description are   provided by the video uploader  as a means to describe its content to the general audience. Moreover, every video is associated with a {\it category},  chosen by the uploader from a pre-defined set of options, including: {\it Autos \& Vehicles}, {\it Pets \& Animals}, {\it Entertainment}, {\it Howto \& Style}, {\it Sports}, {\it Gaming}, {\it Education}, {\it Comedy}, etc.  Every video is also associated (by YouTube) to one or more {\it topics}, extracted from Freebase\footnote{\url{http://www.freebase.com}}, a collaborative semantic knowledge database that covers over 30 million topics, ranging from sports (e.g., baseball) to individuals (e.g., Muhammad Ali). 

For each video-content/video-ad, we also crawled the public statistic data~\cite{Figueiredo2014} that is provided on the HTML page identified by the video id.
This data includes aggregated values of the number of views and exposure time that are accounted for by YouTube. For video-ads only, we also collected the daily time series of both popularity measures.  This statistic data is illustrated in Figure \ref{fig:stats}.

We note that, since each video-ad is an independent video on the system,  these global statistics of video-ad popularity include all accesses to the video, regardless of whether it was paired with a video-content (used as a video-ad) or requested directly. 
 We discuss the implications of this for our analysis in Section \ref{sec:pop}. 

\subsection{Overview of our Datasets} \label{sec:data_overview}

\begin{table}[ttt]
\centering
\caption{Summary of our datasets.}
\begin{tabular}{lccc}
\toprule
& Campus & API   & HTML   \\ 
 & Network & & Stats\\
\midrule
\# of unique   video-contents & 58,082 & 47,007 & -\\  
\# of  unique video-ads & 5,667 & 5,052 & 3,871\\  
\# video-ad exhibitions  & 99,658 & - & -\\
\bottomrule 
\end{tabular}
\label{tab:data}
\vspace{-0.2cm}
\end{table}

We ran the TSTAT tool to collect HTTP requests in the campus network from March $24^{th}$  to November $30^{th}$, 2014. 
Our collected dataset includes  114,709 exhibition time requests, out of which 99,658 (86\%) were successfully matched to video-ad requests, following the heuristic presented in Section \ref{sec:localtrace}. Out of those matches, 2,112 (2\%) were conflicts, which were solved as described in the same section.  
 In total, we identified 58,082 unique ids of  video-contents with  which some video-ad was paired. Such video-ads were identified by 5,667 unique ids.   Table \ref{tab:data} ($2^{nd}$ column) summarizes our dataset collected in the campus network. 

We  collected the API and HTML stats datasets on a single day, May  $27^{th}$, 2015. 
A summary of both datasets is also shown in 
Table~\ref{tab:data}  ($3^{rd}$ and $4^{th}$ columns).  We were able to  crawl the metadata associated with 47,007 video-contents  and 5,052 video-ads,  and we successfully retrieved the popularity time series of 3,871 unique video-ads. We were unable to crawl data for all video-contents and video-ads mostly because of either prohibitive privacy settings by the uploaders or video deletions. 
We note that, even though our API and HTML stats datasets were collected after the campus collection was terminated,  we can still study the global popularity of video-ads on YouTube during the same period covered by the campus dataset by trimming the  time series data accordingly.



\begin{figure}[t]
\centering
\subfigure[Video-ad lifetimes]{\includegraphics[scale=1]{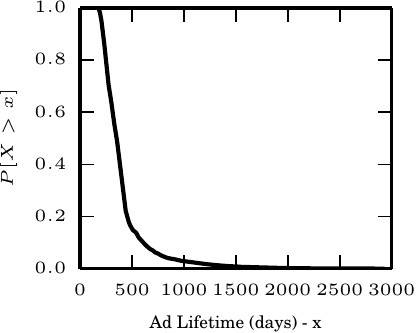}}\hfill
\subfigure[Daily fraction of pairings]{\includegraphics[scale=1]{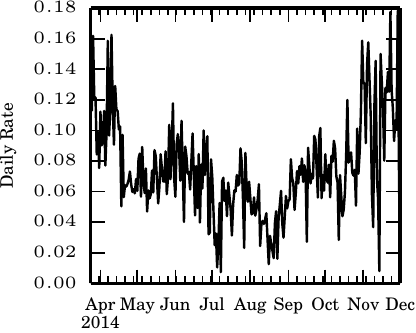}}
\vspace{-0.2cm}
\caption{Overview of video-ads in our datasets. }
\label{fig:vol}
\vspace{-0.2cm}
\end{figure}

Before proceeding, we briefly discuss a few properties of the  video-ads in our datasets. First, we analyze the distribution of their {\it lifetimes}  in the system. The lifetime of a video-ad is defined as the number of days since its upload until our collection of global properties. Figure~\ref{fig:vol}(a) shows the complementary cumulative distribution function (CCDF) of the lifetimes for all video-ads in our API dataset (90\% of all identified video-ads).  Note that all video-ads have been in the system for at least 6 months,  while around half of them have been for more than 1 year. Only a small fraction (6\%) of the video-ads have lifetimes greater than 2 years, though.

Next, we look into the frequency of video-ad to video-content pairings in our campus network dataset. On Figure~\ref{fig:vol}(b) the daily fraction of all video-content requests that are paired with any particular video-ad.
We initially point out that,
on average, the fraction of video-ad pairings is around 7.6\%.  Yet, this fraction increased significantly during the Easter  period (April) and as we approached the holidays of the end of the year (starting from mid October), reaching values from 16\% to 18\%. Thus, in such periods, there is an increase in the expected publicity by a factor of more than 2, when compared to the overall period.  

In the following three sections we present our main findings. The specific dataset used to support each analysis can be inferred based on the information exploited by it, namely, video-ad exhibitions and pairings (campus dataset),  video-ad metadata (API) or video-ad popularity time series (HTML stats).

\section{User Skipping Behavior }
\label{sec:user}
We start our study  by tackling {\it RQ1: How do users consume video-ads?} Recall that YouTube allows users to skip  a video-ad exhibition after a minimum streaming time (usually 5 seconds).  Thus, we answer RQ1 by focusing on the user {\it skipping} behavior, as a step to analyze video-ad exhibition times. 

\begin{figure}[t]
\centering
\subfigure[Duration and time-to-skip]{\includegraphics[scale=1]{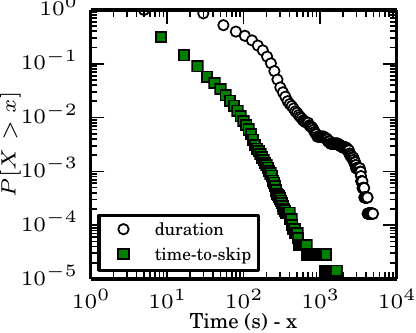}}\hfill
\subfigure[Duration vs. time-to-skip]{\includegraphics[scale=1]{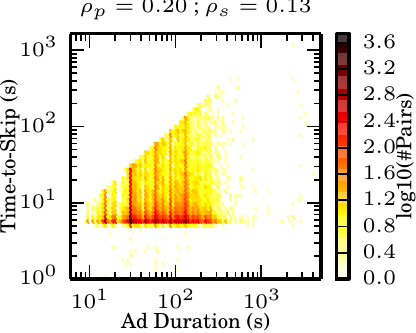}}\hfill
\vspace{-0.2cm}
\caption{User behavior when exposed to video-ads:  duration and time until user skips  exhibition (time-to-skip).}
\label{fig:skip}
\vspace{-0.3cm}
\end{figure}

As a basis for comparison, we first analyze video-ad durations.   Figure \ref{fig:skip}(a), which presents the CCDF of video-ad durations, shows that they  vary greatly across all video-ads.
The mean  is 107 seconds, but the median  is only 60 seconds  and the standard deviation  is 197 seconds.  Moreover, 14\% of the video-ads are very short (below 30 seconds), while  35\%  have durations between 30 and 60 seconds, and 31\%  have  durations above 2 minutes. We also note some rare cases of very long video-ads  (over 1.5 hours)  in our dataset\footnote{Although rare, such video-ads may be exhibited to users since YouTube imposes  no limit on the duration of a video-ad.}. 
 
Next, we analyze the video-ad exhibition times.  The exhibition time is shorter than the duration whenever the user chooses to interrupt and {\it skip}   video-ad exhibition. Thus, we also refer to the video-ad exhibition time as {\it time-to-skip. }
   We first note that  29,442 of the video-ad exhibitions were streamed in full. That is, in 29\% of the video-ad exhibitions, users chose  {\it not} to skip it (despite having the option to do so), watching the video-ad until completion
   The completion rate varies with the category of the video-ad, falling in the range of  17\% (e.g., {\it Music})  to 49\% ({\it Comedy}), but not exceeding 30\% for most categories. 
   We also note that the durations of the video-ads that are exhibited (at least once) in full tend to be somewhat shorter than the overall distribution, as one might expect.
    For example, the average duration of those video-ads is 76 seconds, and the median is only 36 seconds. Also, only 18\% of them have duration above 2 minutes.

The observed fraction of video-ad exhibitions that were streamed until completion contrasts to results in \cite{Krishnan2013}, which reports video-ad completion rates ranging  from 44\% to 95\%.  However, unlike YouTube, the applications analyzed in that work did not allow video-ad skipping. It is interesting to note that a completion rate of 29\% (as in our dataset) is orders of magnitude larger than the click through rates (CTR) often observed in traditional advertising (e.g., 0.01\%) \cite{Schneider2009}.  Such higher video-ad completion rate, particularly in the presence of a skip function,  might suggest a greater user engagement to this new form of online advertising. Yet, such results have to be interpreted in light of two effects. Firstly, it is impossible to skip some video-ads, a fact that increases the completion rate. Secondly, clicking on banner ads comes at a cost from the user. Streaming a video-ad is, in contrast, the {\it default effect}\footnote{http://en.wikipedia.org/wiki/Default\_effect\_(psychology)} provided by YouTube. There is no cost, from a user action perspective, to skip the ad. However, there is a cost related to the interest on the ad from the user. This second cost is what makes the study of the {\it skipping} behavior of users interesting, since it explicitly represents an action from the user of loss of interest on continuing to stream the ad. In the rest of this section, we focus on the behavior of users when they do {\it skip} a video-ad exhibition.


Considering only video-ad exhibitions that were skipped by the user,  Figure \ref{fig:skip}(a) also shows the CCDF of the time-to-skip. Note that, in  more than one third (35\%) of the cases, users  skip  the video-ad exhibition in less than 6 seconds (one second above the minimum), whereas in only  25\% of the cases users wait for more than 10 seconds before skipping the video-ad\footnote{The fractions are similar for all categories of video-ads.}. As also shown in Figure \ref{fig:skip}(a), only 1\% of the video-ads have durations below 10 seconds. Thus, users often skip video-ads shortly after they are allowed to, before streaming a large fraction of their content. Indeed,  we found that, on average, a user skips a video-ad after only 20\% of its content has been exhibited (standard deviation of 19\%). Also, in 50\% of the cases, the skipping is done even earlier,  after only 16\% of the video-ad has been streamed.



We further analyze the skipping behavior by presenting, in Figure \ref{fig:skip}(b), a scatter plot correlating both video-ad duration and time-to-skip.   Each point in the figure is a video-ad exhibition, and  the colors represent the density of points. Only video-ad exhibitions that were skipped by the user before completion are included in the figure. Note that both axes are in log scale. Thus, we computed both the linear Pearson correlation ($\rho_p$) and the Spearman's rank correlation\footnote{A non-parametric measure of statistical dependence between two variables that does not require linear relationships between them.}  ($\rho_s$) between both axes {\it after} taking the logarithm of all values. We found $\rho_p$=$0.2$ and $\rho_s$=$0.13$.  Such low correlations are biased by the  large concentration of points around a time-to-skip (y-axis) of 5 seconds. This concentration implies that many video-ad exhibitions  are largely ignored by the users, who skip them as early as they are allowed to, regardless of their durations.
 

However, there seems to be also another (smaller) group of video-ad exhibitions that are streamed for  time periods roughly proportional to their durations. To uncover this group, we 
focused on video-ad exhibitions that were streamed for much longer than the average, with time-to-skip above the mean ($\mu$=$12.3$) plus two standard deviations ($2 \sigma$=$40$).  
 In those cases, which account for only 2\% of all video-ad exhibitions, the correlations are  indeed much higher ($\rho_p$=$0.57$ and $\rho_s$=$0.50$). Thus, those video-ad exhibition times are roughly proportional to the video-ad durations.
 


 The results from this section may be largely impacted by users that stream a video-ad but do not necessarily watch, or pay attention to, the video-ad. That is, it is impossible to effectively say that users focused their attention to the video-ad being streamed. However, our findings on this section and the rest of the paper reflect an understanding of popularity that is based on ``hits'' and exhibition-times (streaming), similar to how it is accounted for at the server level (e.g., form YouTube) and exploited by video uploaders and marketeers. Thus, our findings provide a view that is perceived by analytics platforms. This factor leads the high correlations between campus views and global views that we shall study in the next section (looking into the popularity properties of video-ads). 


We can summarize our main results on  user skipping behavior as: (1) users often skip video-ad exhibitions as early as they are allowed to, regardless of the video-ad duration, and, on average, 20\% after their beginning; (2) a small fraction of video-ad exhibitions are streamed for a time proportional to their duration; and (3) despite this general trend towards skipping the video-ad, a considerable  fraction of  all video-ad exhibitions  are streamed in full.

\section{Video-Ad Popularity}
\label{sec:pop}

\begin{figure*}[t]
\centering
\subfigure[Local (campus network)]{\includegraphics[scale=1]{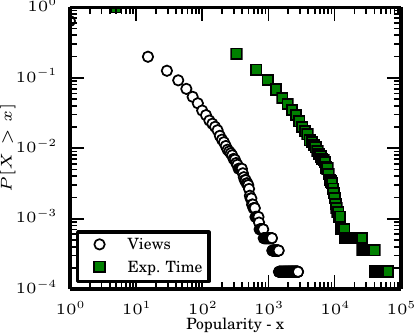}}\hfill
\subfigure[Global (YouTube)]{\includegraphics[scale=1]{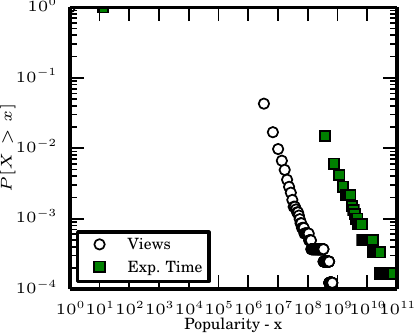}}\hfill
\subfigure[Global filtered (same period of local dataset)]{\includegraphics[scale=1]{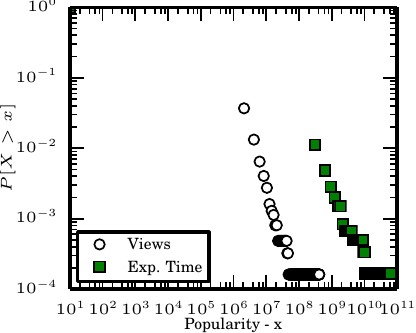}}\hfill
\subfigure[Local vs. global filtered]{\includegraphics[scale=1]{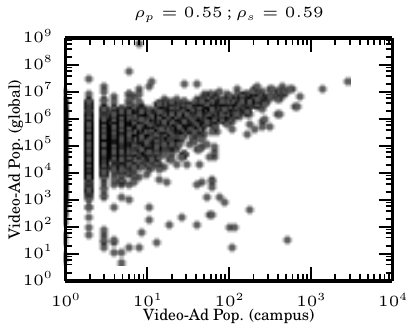}}\hfill
\vspace{-0.2cm}
\caption{Video-ad popularity distributions (in exposure time and number of views) according to different perspectives.}
\label{fig:distributions}
\vspace{-0.2cm}
\end{figure*}

In this section, we address {\it RQ2: How does video-ad popularity evolve over time?} We first
analyze the overall distribution of video-ad popularity (Section 5.1). We then use the daily time series of global popularity of video-ads to analyze the dispersion of popularity temporal evolution and the amount of time until video-ads reach their daily popularity peaks (Section 5.2). Finally,  we use	 a time series clustering algorithm to better understand the different profiles of video-ad popularity evolution (Section 5.3). 


\subsection{ Video-Ad Popularity Distribution}

We analyze the distribution of video-ad popularity using two previously used ad-efficacy metrics, namely,   number of views and    exposure time. The former counts the total number of times the video-ad was exhibited to a user,  regardless of the time of each such exhibition, while  the latter captures the total time during which users were exposed to the video-ad (i.e., total exhibition time).   Our datasets provide two complementary views of each popularity measure: (1) a {\it local} view 	 from the perspective of  the campus network, provided by our traffic logs; (2) a {\it global} view from the perspective of the whole population of YouTube viewers, which  is provided by the API and HTML stats pages (see Section~\ref{sec:data}).
Recall that our API and HTML stat pages represent the popularity evolution of video-ads from the moment the videos were uploaded until the time we crawled  YouTube (May 2015). In order to perform a fair comparison of local and global popularity of video-ads, we filtered our (global) time series data to consider only the popularity gain over the same period covered by our campus dataset (March to November 2014). We refer to this popularity view as {\it global filtered}.

Before proceeding, we emphasize that since each video-ad is itself an independent video on YouTube,  the global popularity of a video-ad   accounts for {\it all} views of the video, 
regardless of whether it was paired with a video-content (exhibited as a video-ad) or accessed as  an independent video.
Thus, even though such global measures of popularity do not necessarily reflect, exactly,  the reach of a video while promoted as a video-ad, they do capture the global interest in its content, and thus may be interpreted as the {\it potential efficacy} of advertisement campaigns that use the video as video-ad.



Figures~\ref{fig:distributions}(a-c)  show the CCDFs of the two  video-ad popularity measures, namely exposure time (in seconds) and number of views, for our three popularity views.  As expected, the popularity measures are much higher when analyzed globally.
Yet, regardless of the perspective and popularity measure, the distributions are highly skewed in nature, following a heavy tail, which is consistent with other studies of video popularity in general  \cite{Figueiredo2014,Cha2009}.  Most video-ads are exhibited only a handful of times and for very short periods, whereas a small fraction of them become very popular.   For instance, only 3\% of the   video-ads were displayed more than 100 times on campus, while only 1.7\% of them had a total (local) exposure time above 1 hour  (Figure~\ref{fig:distributions}(a)). We also found that the most popular video-ad in our campus dataset were also very popular (within the top 0.5\%) in the global and global filtered views. This particular video-ad achieved 2,812 views  and was streamed for 18 hours on campus. In comparison, it received 17,859,680 views and was streamed for 389,653 hours globally during the same time period.  
During its whole lifetime in the system,  the video-ad received 17,947,622  views and was  streamed for  392,239 hours.

We correlated our local popularity measures with the global filtered measures (both in log scale) to gain insights whether our local dataset reflects (to some extent) YouTube's global population in terms of video-ad popularity.  This correlation is shown Figure~\ref{fig:distributions}(d) for popularity estimated by number of views (note the log scale on both axes). Results for exposure time are similar (omitted). We found a Spearman's rank correlation $\rho_s$ of 0.59 (0.54 for exposure time).  Such {\it moderate-to-strong} correlation suggests that, to a reasonable extent, our campus trace reflects the global properties of video-ad popularity on YouTube. This is an interesting result given that YouTube currently receives millions of daily  viewers, whereas our local trace was collected from a campus network whose population includes only tens of thousands of users, most of whom are {\it not} likely to access YouTube every single day.

So far we have analyzed only the total popularity achieved by each video-ad. We are yet to discuss how this popularity evolved over time. Take  the video-ad shown in Figure~\ref{fig:stats} as an example. Although it is one of the most popular video-ad in our
datasets,  most of its popularity is concentrated in a few weeks (based on the time series shown in the figure).
Understanding how video-ad popularity evolves over time can benefit both  content producers, which share a profit of the video-ad's campaign when ads are paired with their content, and content providers.  For example, knowing whether the popularity of a video-ad will be concentrated on a few days or remain popular and generate revenue for longer time periods can ultimately be used to drive monetization strategies as well as caching applications~\cite{Amarie2014,Amarie2014a}.
Thus, in the next two sections, we turn our attention to  how the popularity of video-ads evolves over time.

\subsection{Popularity Dispersion} \label{sec:dispersion}

To study the temporal evolution of video-ad popularity, we used the daily time series of exposure time and number of views crawled from YouTube (global view of popularity). We did not explore our  campus dataset as  it provides only a limited view on popularity evolution. That is,  we found that no video-ad  was exhibited on more than 10 days on campus. Moreover, by using the time series extracted from YouTube, we are able to analyze popularity evolution from the upload of the video-ad until the crawling time.  Specifically,  we address the following questions  in this section: (1) How bursty is video-ad popularity evolution? (2) How much time does it take for a video-ad to reach its daily peak of popularity? 
We focus our  discussion only  on popularity in terms of number of views because very similar results were obtained for both popularity measures. Indeed,  the
  correlations between both time series for each individual video-ad are quite strong  (Pearson correlation $\rho_p$= 0.99 and Spearman
correlation $\rho_s$=0.96, on average), indicating great similarities between them (apart from scale differences). 

\begin{figure}[t]
\centering
\subfigure[Gini coefficient (dispersion)]{\includegraphics[scale=1]{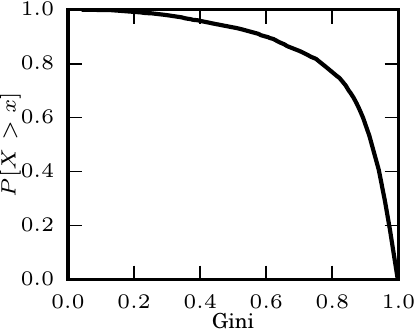}}\hfill
\subfigure[Time to peak]{\includegraphics[scale=1]{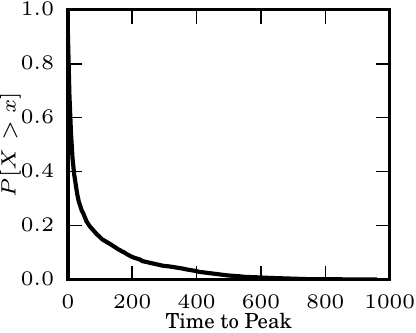}}\hfill
\vspace{-0.2cm}
\caption{Video-ad popularity temporal evolution. }
\label{fig:gini}
\vspace{-0.2cm}
\end{figure}

To answer the first question, we employed a dispersion measure of inequality called Gini score~\cite{Wasserman2010}. The Gini score can be used to measure how bursty a given time series is. Its value ranges from 0, when the total popularity acquired by a video-ad is roughly homogeneously dispersed over its lifetime, to 1, when the popularity is concentrated on a single day.  According to Figure~\ref{fig:gini}(a), which  shows the CCDF of the Gini scores computed for the video-ads in our dataset,   84\% of the time series have a score higher than 0.7, and 57\% have a score higher than 0.9.  Thus,  most video-ads have their popularity evolution concentrated on a few days. Yet, we do observe some video-ads with low Gini scores:  4\% of all video-ads have  scores below 0.4, suggesting that they succeeded in attracting attention for longer time periods. 
One might wonder whether there is a  correlation between the video-ad lifetime and its Gini score (e.g., whether video-ads that have been more recently uploaded have lower Gini scores).  However, we found no clear trend between video-ad lifetime and Gini score (Pearson  $\rho_p$ = -0.27 and Spearman  $\rho_s$ =  -0.2) in our dataset. 

To tackle the second question,  Figure \ref{fig:gini}(b) shows the distribution of  time (in days) from the video-ad upload  until its daily popularity peak\footnote{In case of ties -- multiple days with the same popularity peak -- we took the first day.}. Typically, most video-ads (69\%) reach their popularity peak within one month after upload, while for half of them the peak occurs in at most 12 days after upload.  Thus, video-ads often peak in popularity very early in their lifetimes, possibly as a reflection of advertisement campaigns that are initiated shortly after the upload.  However, this is not always the case. For  example, for 10\% of the video-ads, the popularity peak occurred only after 6 months since upload\footnote{Those might be videos that were first uploaded to the system and only used in video-ad campaigns much later.}. On average, the number of days  until popularity peak is 56.  If we normalize the time-to-peak by the video-ad lifetime, we observe that, on average, a video-ad takes only 12\% of its lifetime to reach its daily popularity peak (median of 4\%).



\begin{figure*}[t]
\centering
\subfigure[C1]{\includegraphics[scale=.145]{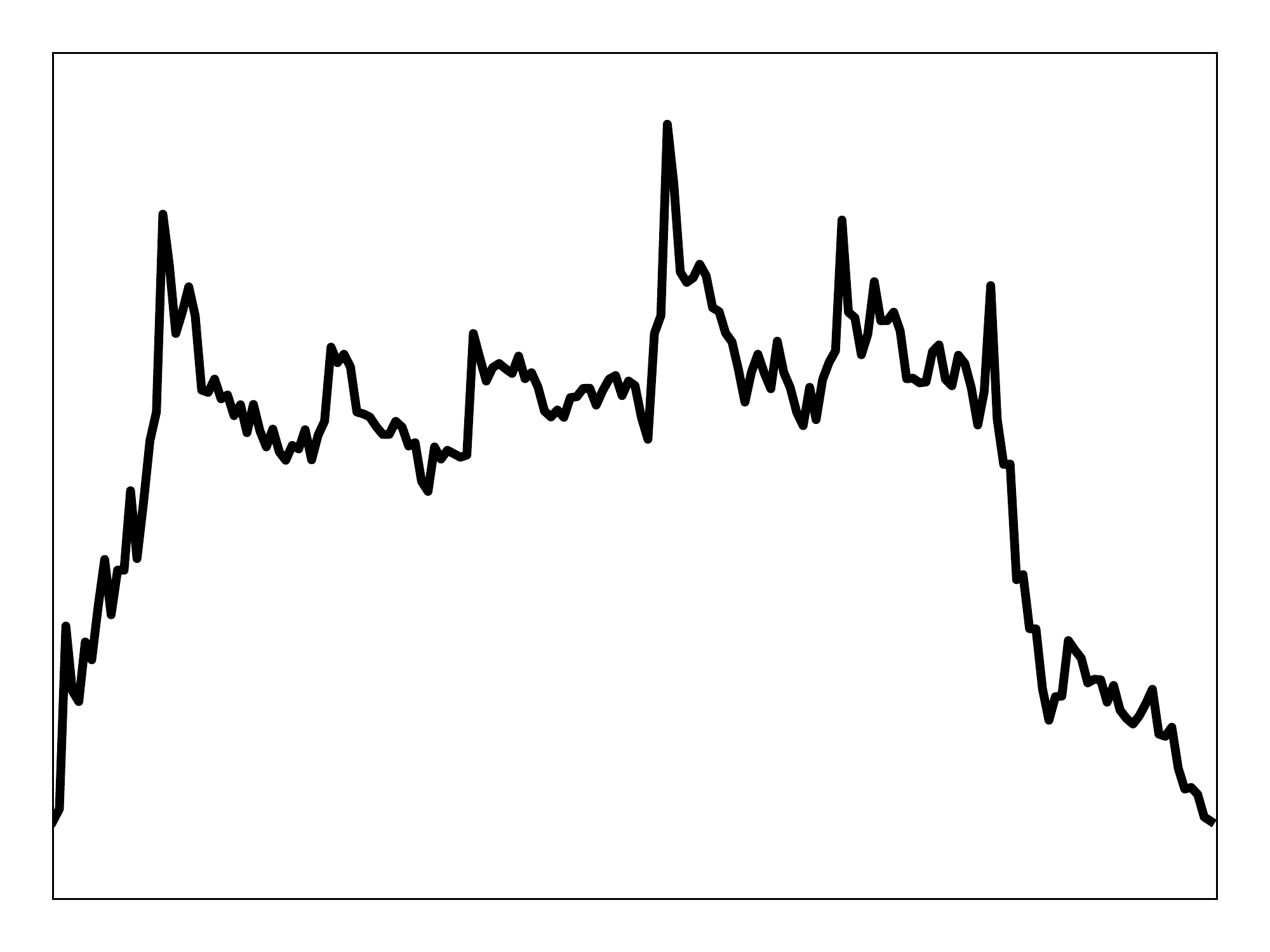}}\hfill
\subfigure[C2]{\includegraphics[scale=.145]{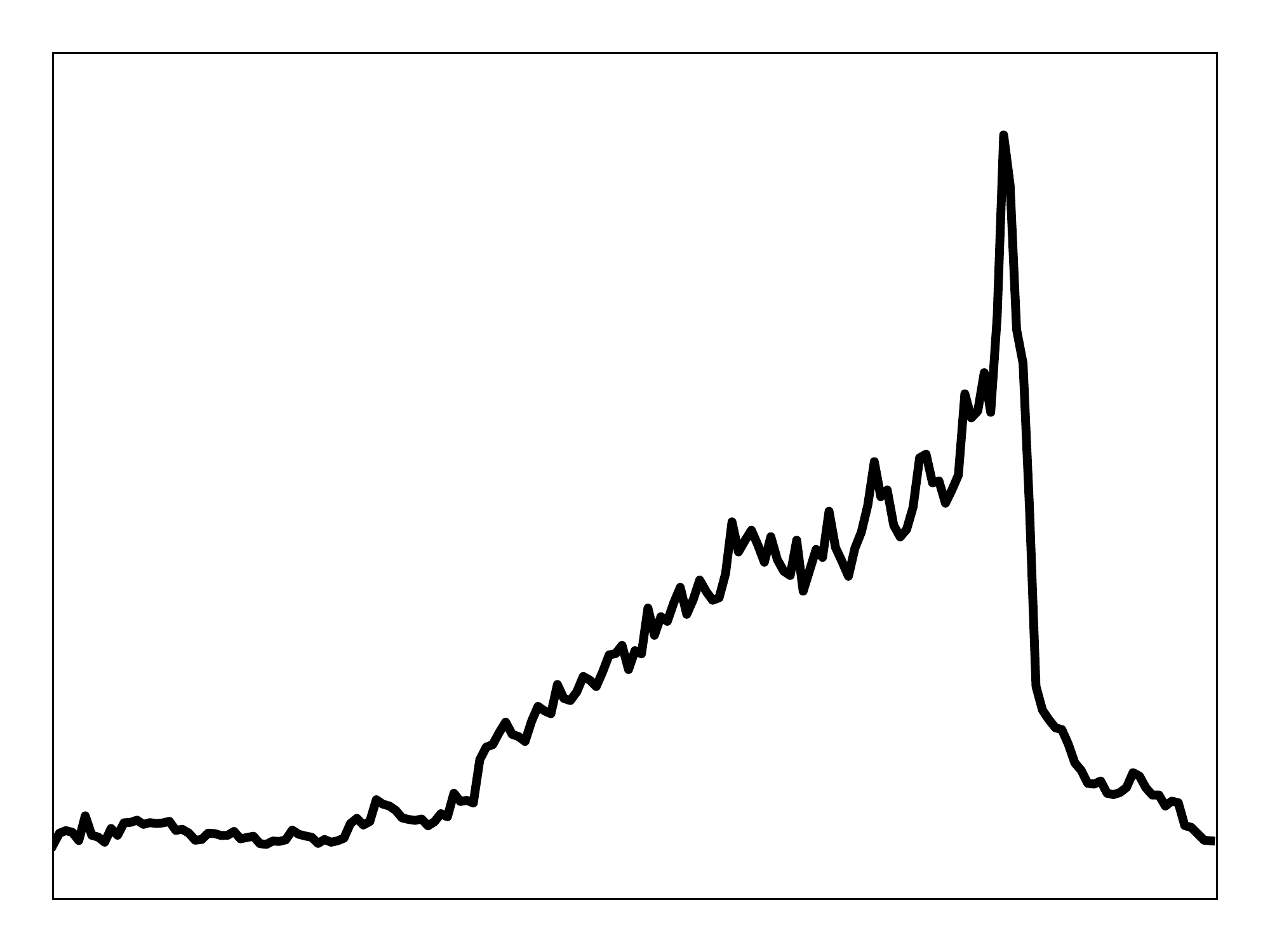}}\hfill
\subfigure[C3]{\includegraphics[scale=.145]{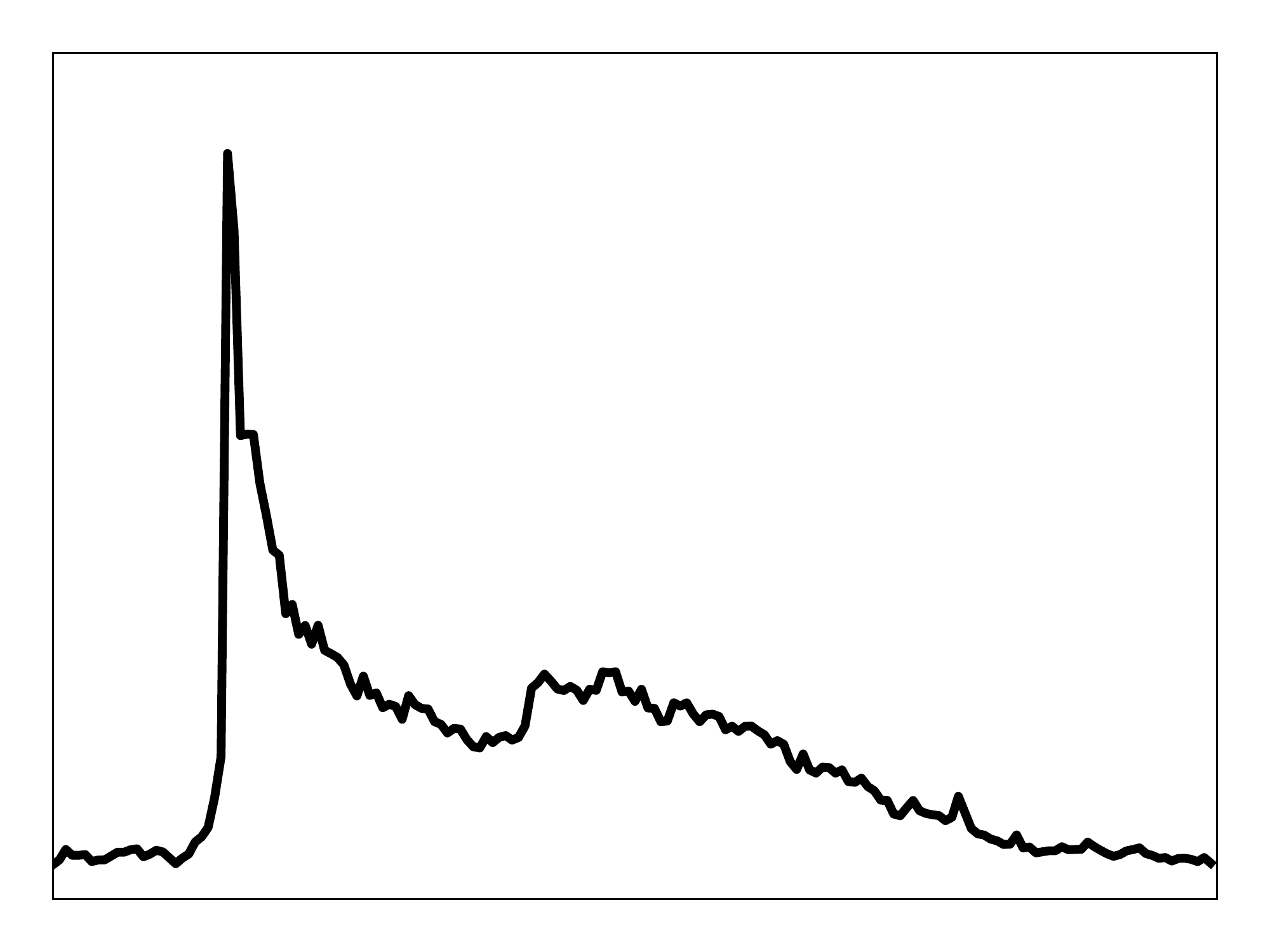}}\hfill
\subfigure[C4]{\includegraphics[scale=.145]{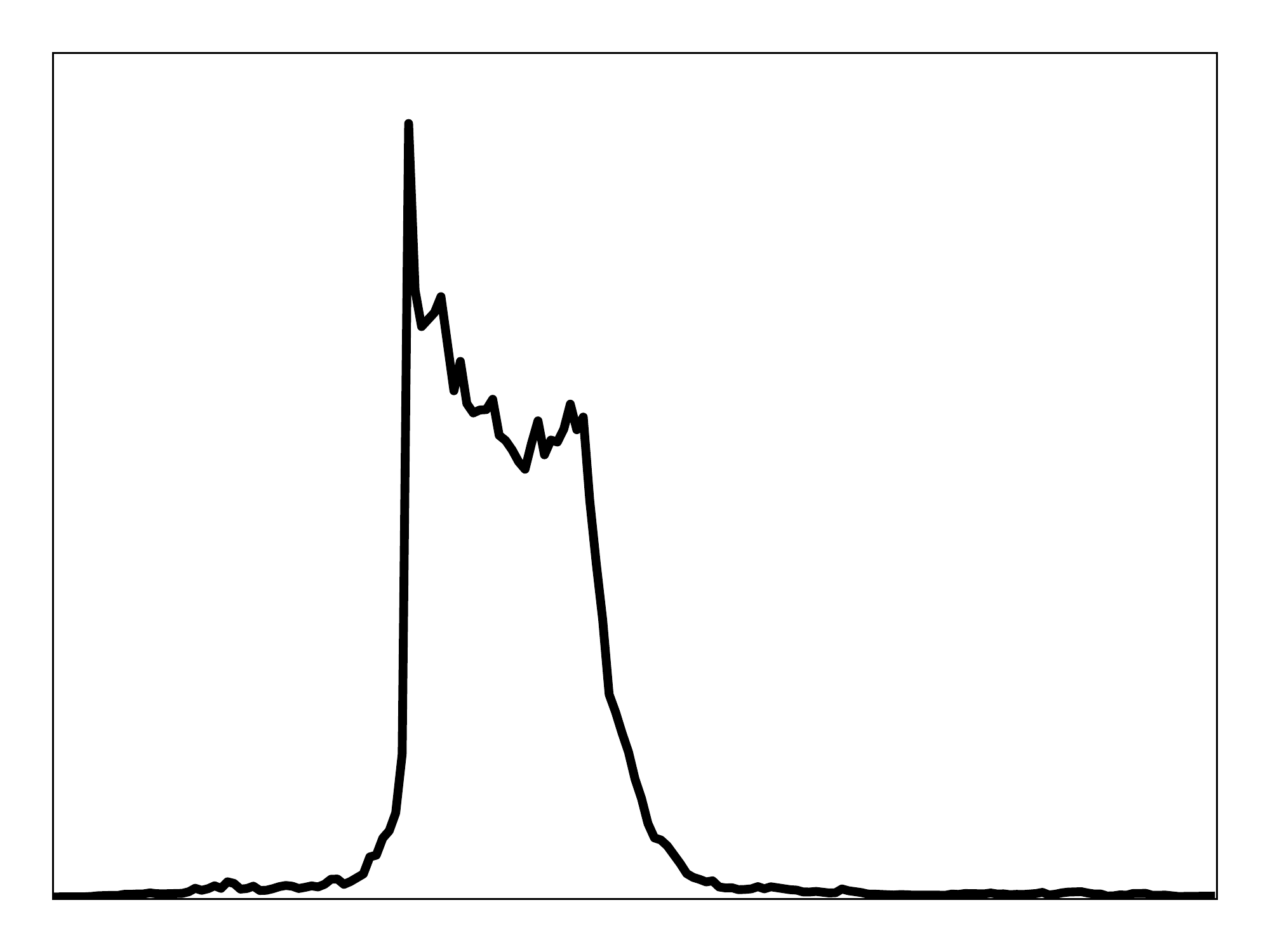}}\hfill
\subfigure[C5]{\includegraphics[scale=.145]{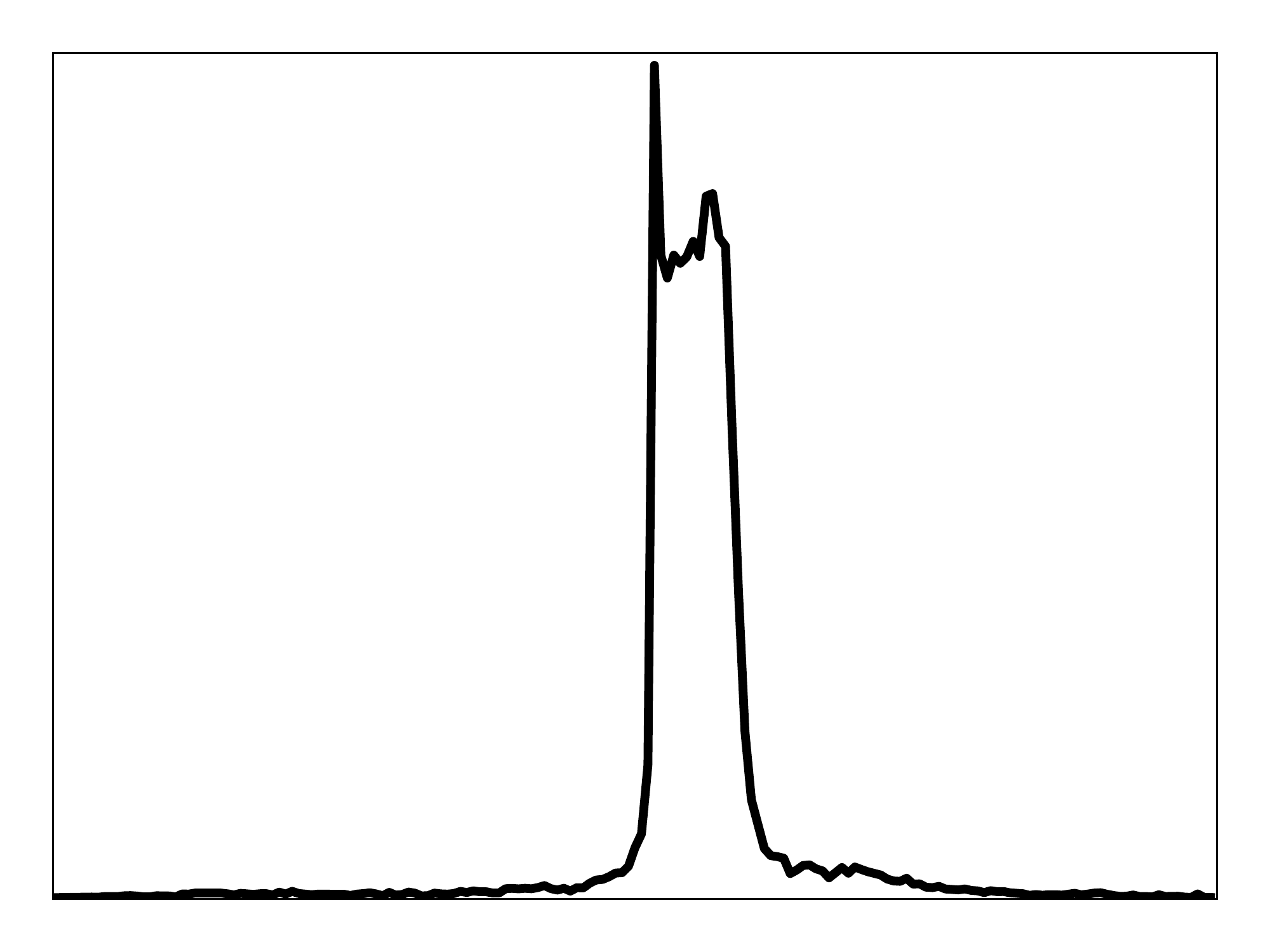}}\hfill
\subfigure[C6]{\includegraphics[scale=.145]{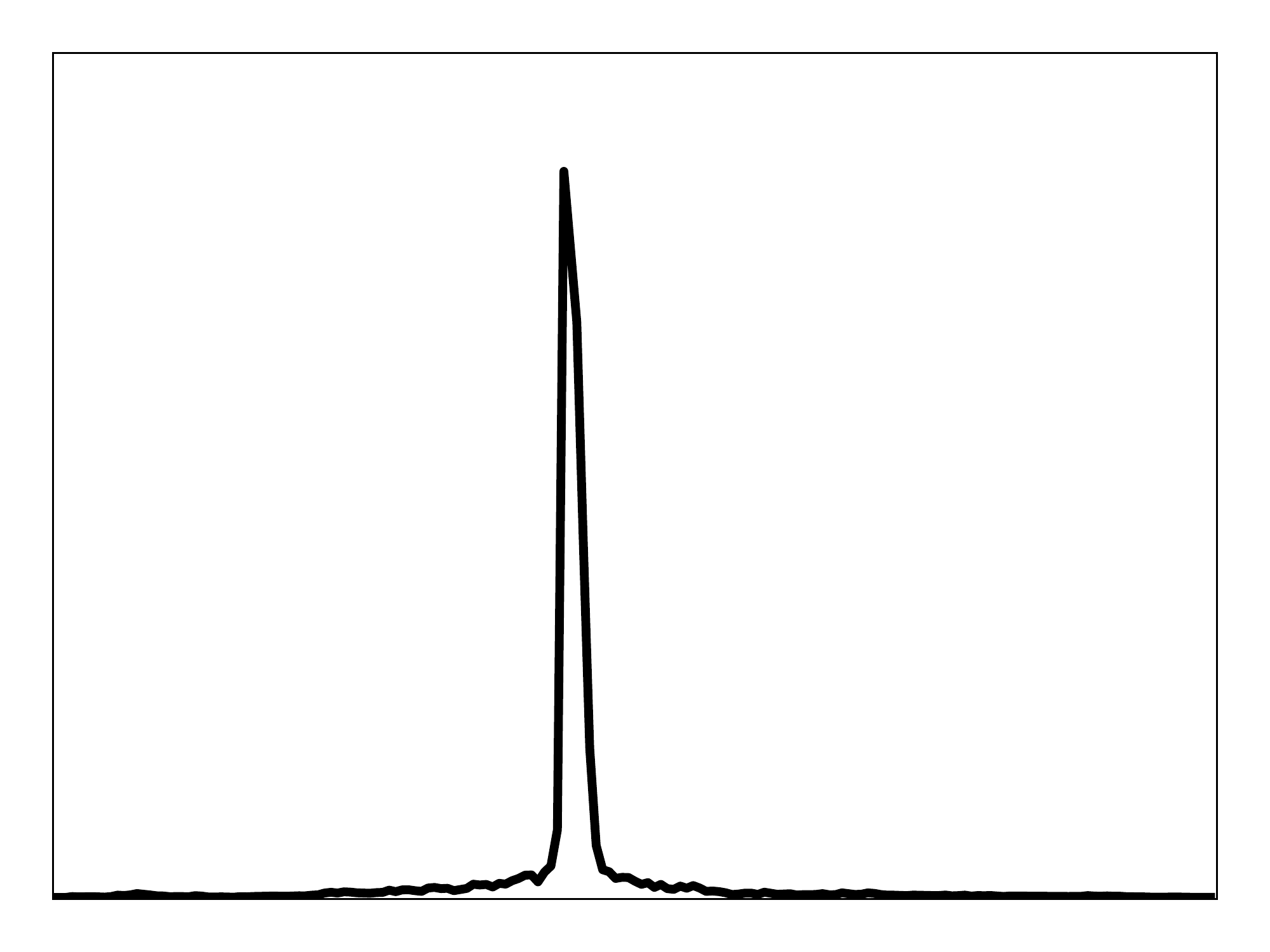}}
\vspace{-11pt}
\caption{Trends (cluster centroids) of video-ad popularity evolution over time.}
\label{fig:clusters}
\vspace{-11pt}
\end{figure*}

\begin{table*}[t]
\footnotesize
\centering
\caption{Properties of each trend (cluster) of video-ad popularity evolution.}
\begin{tabular}{lcccccc}
\toprule
& C1 & C2 & C3 & C4 & C5 & C6 \\
\midrule
\# video-ads & 69 & 108 & 109 & 293 & 467 & 569\\
Average Number of Views & 1,486,175 & 1,869,906 & 4,882,094 & 1,789,798 & 1,451,894 & 984,175\\
Average Exposure Time & 203,640,554 & 159,293,660 & 629,686,649 & 99,386,939 & 81,300,652 & 60,885,487\\
Average Exposure Time / Number of Views & 137.02 & 85.19 & 128.98 & 55.53 & 56.0 & 61.86\\
Average Gini & 0.24 & 0.61 & 0.58 & 0.82 & 0.9 & 0.92\\
Average Time to Peak & 66 & 69 & 37 & 25 & 20 & 14\\
\bottomrule
\end{tabular}
\label{tab:clusterdesc}
\end{table*}

Next, we deepen our investigation of video-ad popularity  by identifying common profiles (trends) of popularity temporal evolution.


\subsection{Profiles of Popularity Evolution}

Towards identifying profiles of popularity temporal evolution of video-ads, we made use of a time series clustering algorithm called K-Spectral Clustering (KSC)~\cite{Yang2011}, which has been successfully used to study the patterns of popularity dynamics of social media content  \cite{Figueiredo2014,Yang2011}.  KSC is a K-Means based algorithm that groups different time series into clusters based simply on the {\it shape} of the curves. It does so by using a distance (or similarity) metric that respects scale and time shifting invariants. That is, two video-ads that have their popularity dynamics evolving according to similar processes will be assigned to the same cluster by KSC, regardless of the popularity values. For example, two time series that are stable over time except for a  peak in a day will be grouped together, regardless of when the peak occurred (time shifting invariant) and the peak value (scale invariant). 
 By taking into account both of these invariants, we can focus on the overall {\it shapes},  or  {\it trends}, that define the governing properties of popularity temporal evolution of video-ads. These trends are represented by the cluster centroids (or averaged time series)  produced by  the KSC algorithm.


KSC requires that all time series have the same length.  Thus, we trimmed our video-ad popularity time series to include only the first 180 days. Recall that, as discussed in Section \ref{sec:data_overview},  all video-ads in our dataset have been in the system for at least 180 days.
We note that such trimmed time series do include the daily popularity peaks for most  video-ads:  the peak occurs within the first 180 days after upload for 90\% of the video-ads (see Section \ref{sec:dispersion}). 
  Moreover, for the sake of a fair comparison between the identified profiles, we focused our analysis on video-ads that attracted at least a minimum of 180,000 views (1,000 daily views {\it on average}).  In total, we clustered 1,615 video-ad time series that meet this criterion.


The KSC algorithm also requires the choice of a number $k$ of clusters.  We employed  various clustering quality measures suggested by previous authors (coefficient of variation, silhouette and clustering cost based scores)~\cite{Figueiredo2014,Yang2011} to choose this value. We also performed a visual inspection of the cluster centroids and  individual cluster members for different values of $k$. In a few  cases, we manually merged clusters that, despite being identified as separate groups according to some clustering quality measure, did contain members with very similar popularity evolution patterns. 

Based on all these heuristics, we identified  $k=6$ clusters.  
The cluster centroids are shown in Figure \ref{fig:clusters}. Each centroid corresponds to an ``average'' popularity curve for the video-ads in the cluster, capturing, in general terms, the popularity dynamics of the individual members of the respective cluster. Scales on both axes are omitted to emphasize the scale and time shifting invariants.   Table \ref{tab:clusterdesc} summarizes the characteristics of  each cluster by presenting the number of video-ads, as well as the average values of exposure time, number of views, ratio of exposure time to number of views, Gini score and time to peak  of the members of each cluster.


Cluster C1 (Figure \ref{fig:clusters}(a)) consists of video-ads that have succeeded in attracting user attention over a larger number of consecutive days. Indeed, this cluster has the smallest average Gini score (Table \ref{tab:clusterdesc}). 
Clusters C2 and C3 (Figures \ref{fig:clusters}(b-c)), in turn, exhibit complementary popularity trends: video-ads in C2 tend to have a slow growth of  popularity  followed by a sharp decay, while video-ads in C3 exhibit a sharp initial growth of popularity followed by a slow decay. 
Note that, consistently with such trends, video-ads in C2 take more than twice longer than video-ads in C3 to reach their popularity peak.
These trends are interesting since similar patterns of growth and decay have previously been accounted for as viral-like propagation over social networks~\cite{Yang2011}.  Their average Gini scores are similar (around 0.6) but higher than that of C1. Thus, these video-ads tend to concentrate their popularity in fewer days.


The popularity trends captured by clusters C4-C6 (Figures \ref{fig:clusters}(d-f))  exhibit sharp increase and decrease of popularity. Also,  video-ads in these clusters remain popular for much shorter time periods, compared to those in C1-C3 (note the higher Gini scores).  The main distinguishing feature of C4-C6 is  the time window during which the video-ad attracted user attention, which is longer in C4 and shorter in C6.  These patterns may reflect advertisement campaigns having different durations. Some video-ads are publicized for a few weeks, others for only a few days. Note that the time to peak tends to decrease with the concentration of popularity, suggesting  that less dispersed clusters tend to  peak in popularity earlier.

  Overall, video-ads in C1-C3 tend to attract more user attention in both total number of views and total exposure time, at least on average.  This seems to suggest that ad-campaigns that manage to remain attractive for longer time periods will eventually become the most popular ads, which is somewhat expected. Yet,  not all video-ads can remain attractive for long periods. For instance, seasonal ad campaigns,  such as those related to Christmas, face the challenge of attracting a lot of  attention over  short time windows.  
  
  Moreover,   video-ads in C1-C3 have also higher ratio of exposure time to number of views (Table \ref{tab:clusterdesc}), implying that they tend to attract more attention of {\it individual viewers} as well. Looking at some of the most popular video-ads in clusters C1-C2, we found two musical clips. These two videos, which were also publicized as video-ads, will likely attract viewers regardless of the ad-campaigns they are used in. Thus, one interesting direction of future work is to  analyze the importance of video-ad campaigns to the  ultimate growth of popularity achieved by a video.
 
  In order to understand the nature of the video-ads in each of these clusters, we looked into their video categories (e.g., Music, Pets, Entertainment etc.). The clusters C1-C2 have the majority of their video-ads as members of the {\it Music category}. This fact can explain why the attention received by video-ads in these clusters extend for longer periods of time. Previous research has also found the effect that music videos remain attracting attention over time~\cite{Figueiredo2014}. The clusters C3-C6 presented {\it Entertainment} as the most popular category. We believe that this is a category of broad semantics (covers various topics) that is exploited by advertisers of products/goods which cannot be described by other YouTube categories. Nevertheless, these ads fail to attract attention over long periods of time, and thus exhibit rise-and-fall dynamics~\cite{Figueiredo2014}.

\section{Video-Content to Video-Ad Pairs}
\label{sec:match}
We finally turn to {\it RQ3: What are the relationships (if any) between a video-ad and the video-contents with which it is associated?} We address it by measuring the correlation between the popularity of a  video-ad and the popularity of the video-contents with which it was paired (Section~\ref{sec:paircorr}), and  the content similarity  between video-ad and video-content in each  pairing (Section~\ref{sec:pairtxt}). 

\subsection{ Video-Ad and Video-Content Popularity} \label{sec:paircorr}

\begin{figure}[t]
\centering
\subfigure[Exposure Time]{\includegraphics[scale=1]{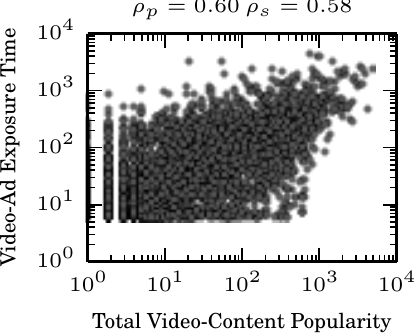}}\hfill
\subfigure[Number of views]{\includegraphics[scale=1]{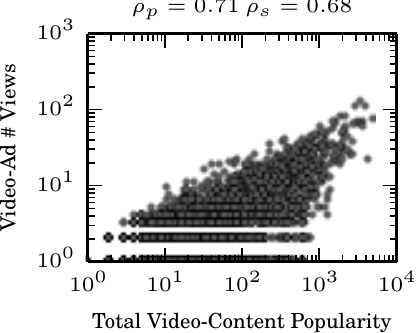}}\hfill
\vspace{-0.2cm}
\caption{Popularity of video-ad versus total popularity (in \# views) of all video-contents that were paired with the video-ad  (measured in the campus network).}
\label{fig:corrpop}
\vspace{-0.3cm}
\end{figure}

We measured the correlation between video-ad popularity and video-content popularity as follows.
 For each video-ad in our campus dataset, we  summed the  total popularity, captured by the number of views, of {\it all} video-contents that were paired at least once with the given video-ad in the dataset. Note that this sum includes all requests to those video-contents  in our dataset (even when they were not paired with the given video-ad). Figures~\ref{fig:corrpop}(a-b)  show the correlations of  this value with the two previously defined video-ad popularity measures, namely, exposure time and total number of views (both axes in log scale).
We focus only on popularity measures computed inside the campus, as we do not know all  pairings involving a particular video-ad  from the global data collected.  



 Figure~\ref{fig:corrpop} shows  reasonably strong linear correlations between the  popularity of the video-ad and the total popularity of all video-contents with which the ad was paired. The Pearson correlation ($\rho_p$) ranges from 0.6 (when correlating with  exposure time) to 0.71 (when correlating with the number of views). Similarly, the Spearman's rank correlation  ($\rho_s$) ranges from 0.58 to 0.68. Such strong correlations are intuitive: they suggest that  the traffic to popular video-ads will be driven, to a large extent, by the aggregated popularity of all video-contents these ads are paired with. More popular video-contents create more opportunities for video-ads to grow in popularity. Thus, advertisement campaigns have a higher chance of being more successful when video-ads are matched to contents that are currently popular, or will grow/remain popular over time.

 However,  the correlations are  weaker when considering individual video-contents, possibly due to the heterogeneity of the video-contents with which the same video-ad is paired. For example, when considering the average popularity of the video-contents, $\rho_p$  ranges from 0.32 to 0.34,  and  $\rho_s$ is equal to 0.36 (for both measures).  Yet, the correlation between video-ad popularity and the total number of videos with which the ad was paired is quite strong ($\rho_p$ and $\rho_s$ exceed 0.83).  Pairing a video-ad with more video-contents raises the chance of hitting a content that will be very popular,  thus increasing the probability of the video-ad inheriting its audience and becoming popular as well.



These strong correlations suggest that effective content popularity prediction methods might be exploited in the design of ad-to-content pairing approaches, aiming at maximizing video-ad popularity. Indeed, content popularity prediction has recently gained a lot of attention \cite{Radinsky2013,Pinto2013}, often driven by the goal of designing more effective advertising services. Yet, no prior work has analyzed the correlations  between video-content popularity (the target of the predictions) and video-ad popularity,  thus offering quantitative results to support such goal, as we do here.


\subsection{Content Similarity} \label{sec:pairtxt}

\begin{figure}[t]
\centering
\subfigure[TF weights]{\includegraphics[scale=1]{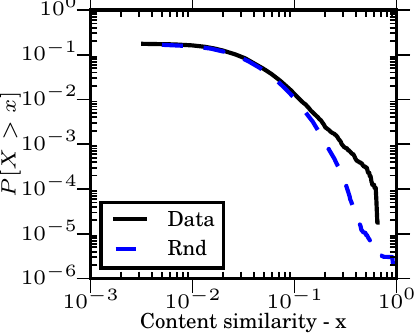}}\hfill
\subfigure[IDF weigths]{\includegraphics[scale=1]{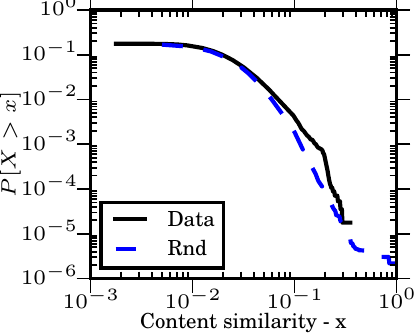}}\hfill
\vspace{-0.2cm}
\caption{Content similarity between video-ad and video-content  in the real (Data) and random (Rnd) datasets.}
\label{fig:simccdf}
\vspace{-0.3cm}
\end{figure}

We now quantify the content similarity between the video-ad and each  video-content with which it was paired. To that end, we use the category,  list of  Freebase topics, title and description associated with each video (content and ad), crawled from the YouTube API. 

  Initially, we quantified the fraction of video-ad to video-content pairings in which both videos have the same YouTube category. We found that such fraction is very low (9\%).
Similarly,  the fraction of pairings in which both videos have at least one Freebase topic in common is also very small (1\%). Freebase topics are more specific than a category, and capture the semantics of a video as determined by the YouTube platform. This lack of similarity proves evidence that the videos in most pairings may have quite different semantic contents, as we further investigate next.



We  then turned to the title and description features of each video to build a textual representation of the video's content.
Specifically, we  first
pre-processed the title and description features   by:  (1) combining the contents of both strings; (2) de-capitalizing the words; (3) removing accents, punctuation, and stop-words\footnote{Stop-words refer to the most common words in a language. We removed stop-words in both English and Portuguese (e.g., {\it an, or}).}; (4) removing words that appear only in the representation of some video-ad (but no video-content) or only in some video-content (but no video-ad).

The content of each video $v$  was then represented as  a bag of words $T_v$. Let  $\mathcal{T}$ be the set of  bags of words representing all videos  (ads and contents) in our dataset, and $\mathcal{V}$ be the vocabulary size  (i.e., total number of unique words) of  $\mathcal{T}$. 
 Each bag $T_v$  can thus be mapped to a vector  $\mathbf{t}_v$ where each entry of the vector corresponds to a word $i$ in $\mathcal{V}$, that is: $ \mathbf{t}_v = < w_{T_v,1}, w_{T_v,2}, \cdots, w_{T_v,|\mathcal{V}|}>. $
 
 
We experimented with four heuristics as weighting factors $w_{T_v,i}$. The {\bf binary} heuristic, although simple, fails to capture the {\it descriptive} and {\it discriminative} properties of the words in $T_v$.  We capture the descriptive strength of a word \ using the {\bf Term-Frequency (TF)} heuristic. We use the {\bf Inverse Document Frequency (IDF)} heuristic to estimate the discriminative capacity of a word and we combined both descriptive and discriminative capacities by taking the product of both metrics, using the {\bf TF*IDF} heuristic.

Given two vectors $\mathbf{t}_a$ and $\mathbf{t}_c$ representing a video-ad and a video-content with which it was paired, we estimate the content similarity between both videos by the cosine of the corresponding vectors (using each weighting heuristic).
The cosine varies from 0, when the textual representations of both videos share no common words, to 1, when they are equal.

As baseline for comparison, we also built 500 random datasets of video-ad to video-content pairings. Each random dataset was created by taking the pairings in our real dataset and   randomly shuffling the ids of the video-ad and video-content in each pair. 


Figures~\ref{fig:simccdf}(a-b) show the CCDFs of measured similarities in our real dataset (Data) and in the 500 random datasets (Rnd), for two of the four weighting heuristics. For each heuristic, we compared the two distributions by testing whether the measured similarities are greater than the similarities in the random datasets (i.e., above random chance).
To that end, we applied two  non-parametric statistical tests, namely one-sided Kolmogorov-Smirnov  and one-sided Mann-Whitney-U.  According to both tests, the similarities in our dataset are greater than the similarities in the random datasets ($p$-value < 0.05). Yet, 
as shown in Figures~\ref{fig:simccdf}, in practice the two distributions are very similar (with differences coming up mostly in the tail). Moreover, similarity values are often very small:  the median is 0 and the mean is below 0.01 in both real and random datasets, regardless
of the weighting heuristic used. Similarly, the $90^{th}$ percentiles of the distributions do not exceed 0.04, again in both real and random datasets. These results provide evidence that most often video-ads are not paired with video-contents of similar semantic content (as captured by their title and description).

While similarities tend to be low, there is still a chance that pairings with higher similarities tend to lead to more popular video-ads. That is, users may show more interest in video-ads that are paired with similar video-contents. 
We investigated whether this is true in our dataset by measuring the correlation between the popularity of a video-ad and the average cosine similarity of all the pairings involving the video-ad (both in log scale). We found that these correlations are reasonably low. That is Spearman and Pearson coefficients of at most 0.33 for all four heuristics. This result indicates that popularity is not explained by similarity, as was the case when correlating video-ad and video-content popularity.   

In sum, our results indicate that video-ad to video-content pairings are, in most cases, dissimilar in terms of textual content. We also found only weak evidence  that more similar pairings tend to lead to more popular video-ads.  One question that arises  then is whether one can design novel targeted  advertising techniques that, by taking the  similarity between video-ads and video-contents into account when pairing them, lead to more successful (popular) video-ads. This is a subject for future work.

\section{Discussion and Future Work}
\label{sec:conclusions}

Social media applications rely heavily on their audience to generate revenue. Content providers (i.e., the application) should aim at offering an enjoyable experience to their audience, while still relying  on content producers  to  attract users, and on online advertisers to build ad campaigns upon which all parties can profit. For example, on YouTube, advertisers usually pay the application for every 1,000 video-ad streams, while content producers receive profit for every 1,000 views of video-ads that were paired with their content.  Understanding the factors behind the success of an ad campaign in such complex  ecosystem is quite challenging, but it is also  key to the design of more effective and profitable advertising strategies.

  In this paper, we took a step towards building such understanding by shedding light into how  one particular type of online ad,  video-ads,  are currently  consumed on YouTube.   Driven by three  research questions, we   presented a thorough measurement study covering aspects related to how users individually respond to video-ad exhibitions, different profiles of video-ad popularity evolution,  the role of video-contents to attract popularity to video-ads, and  the extent to which the current ad-to-content pairing strategy employed on YouTube exploits the semantic similarity between the videos.


  Our study revealed that, even though YouTube users often skip video-ad exhibitions as early as possible, the fraction of  exhibitions that are streamed until completion is reasonably high (29\%).  If compared to the click through rates of traditional advertising (often below 0.01\%), this result might suggest a greater user engagement and thus a potentially more effective means of online advertising. Yet, this result should   be taken with caution.  It is important to consider that watching a video-ad in full is the {\it default effect} provided by YouTube. The default effect in traditional click advertising is  {\it not to click} on the ad, which may have a role on its lower efficacy.  While our  work offers a first analysis of user engagement to YouTube video-ads,  follow-up studies, possibly including experiments with volunteers, should be performed to 
  compare the effectiveness of both strategies in light of default effects. Such user experiments, along with the results we present here, would provide  a broader view of the user experience, which, in turn, could offer valuable insights into the design of advertising strategies that entertain the users, while still generating profits to the other parties.

We also found that,  although most video-ads have their popularity concentrated on a few days,  some of them remain popular for much longer. Indeed, our study uncovered six different profiles of video-ad popularity evolution.  In light of such profiles, one question that arises is: {\it What is the most effective means to pair video-ads and video-contents so as to increase the chance of the video-ad remaining popular for longer periods?} 
Content producers would be interested in attracting video-ads that  remain popular for as long as possible (e.g, video-ads in clusters C1-C3) to maximize revenues.  Advertisers, in turn, are interested in pairing their video-ads with contents that will lead users to their products. As we have shown, there is a trend towards video-ads that are paired with popular contents (and a larger number of video-contents) inheriting such viewers and becoming popular as well.    Yet, our study also revealed that video-ad to video-content pairings are still mostly dissimilar in terms of content similarity (as captured by the textual features of both videos).   This result motivates future investigations on whether  contextual advertising strategies can be more effective in generating revenues for both parties. 

Finally, the results uncovered in this paper have focused on user behavior, popularity properties and contextual advertising. One important factor~\cite{Gill2013} that we are currently exploring as future work is on the nature of targeted (personalized to the users demographic) ads. 
Targeted ads account for a large fraction of online advertising nowadays, and is the focus of studies of different ad-auction strategies~\cite{Gill2013,Liu2014}.  Nevertheless, the results we uncovered in this study can also be exploited by different ad-auction strategies~\cite{Gill2013,Liu2014}. For instance, the correlations between video-content popularity and ad-popularity can be used to estimate the exhibition time of ads. Premium video-content which attracts more exhibition to ads can exploit higher prices in ad auction bids. While in contrast, the lack of correlation between the similarity of video-content and ads with popularity, indicates that this factor will likelly not lead to more viewers, and thus should not affect bidding prices. 



\section*{Acknowledgments}
This work was partially supported by the project FAPEMIG-PRONEX-MASWeb,
Models, Algorithms and Systems for the Web, process number APQ-01400-14, and by individual grants from CNPq, CAPES, and Fapemig. We also thank Prof. Nazareno Andrade for the feedback on preliminary versions of this work.

\balance

{
\bibliographystyle{abbrv}
\bibliography{bibs}
}
\end{document}